\documentstyle[12pt]{article}
\newcommand{\doe}{This work was supported by the Director, Office of Energy
                  Research, Division of Nuclear Physics of the Office of High
                  Energy and Nuclear Physics of the U.S. Department of Energy
                  under Contract No. DE-AC03-76SF00098 \\
                    E-mail TOPOR@PADOVA.INFN.IT ; TOPOR@ROIFA.BITNET \\
		    TOPOR@NT1.PHYS.COLUMBIA.EDU}

\begin{document}

\begin{titlepage}

\begin{flushright}
     {\large CU - TP- 1995}\\
      {\large DFPD95/NP-48-1995\\}
  \end{flushright}
\vskip 1\baselineskip
\renewcommand{\thefootnote}{\fnsymbol{footnote}}
\setcounter{footnote}{0}
\begin{center}
\baselineskip=24pt
\mbox{}\\[2ex]
{\Large \bf Nucleus - Air Interactions for Extensive Air Showers Experiments
 at Ultrahigh Energies{\footnote{\doe}}}\\[2ex]

\baselineskip=18pt
{\large \bf V.Topor Pop}\\[2ex]
{\em Physics Department, Columbia University, New York, NY 10027}\\
{\em and}\\
{\em Dipartimento di Fisica "G.Galilei";Istituto di Fisica Nucleare}\\
	{\em Sezione di Padova,}\\
{\em Via Marzolo 8-35131, Padova,Italy}\\[2ex]
 {\large \bf and}\\
{\large \bf H.Rebel}\\[2ex]
{\em  Kernforschungszentrum Karlsruhe,Institut fur Kernphysik III,}\\
    {\em  P.O.Box 3640,D-76021 Karlsruhe,Germany}\\[2ex]
      \mbox{}\\[4ex]

\today\\[2ex]
\vskip 2cm

{ \large \it Work to be submitted to Astroparticle Physics}
\end{center}

\newpage
\vskip 10cm

\begin{abstract}
\normalsize
\baselineskip=24pt

  The HIJING and VENUS models of relativistic hadron-nucleus and
 nucleus-nucleus collisions are used to study interactions of proton and
 nuclei with nitrogen, specific for the extensive air shower
  developments initiated by cosmic rays in the atmosphere.
  The transverse energy,transverse momenta and secondary particles
  produced spectra as well as their energy and mass dependence
  were investigated in detail. Results are presented with
 particular emphasis on the contributions of minijets in
 HIJING model  and validity of superposition models
 in this energy range .
\end{abstract}

\end{titlepage}

\newcommand{\lsim}
{\ \raisebox{2.75pt}{$<$}\hspace{-9.0pt}\raisebox{-2.75pt}{$\sim$}\ }
\newcommand{\gsim}
{\ \raisebox{2.75pt}{$>$}\hspace{-9.0pt}\raisebox{-2.75pt}{$\sim$}\ }

\baselineskip=18pt
\parindent=0.25in
\abovedisplayskip=24pt
\belowdisplayskip=24pt

\section{Introduction}

The investigation of the detailed shape of the energy spectrum and the
 mass composition of primary cosmic rays is currently a most active
  field of astrophysical research \cite{rebel1}.

Experiments on satellites or with balloon-borne detectors give
 information up to energies of ca. $\, 10^{14}$ eV \cite{mul1}.
 Due to their limitations
  in size and weight they can hardly be extended
   beyond$\,\, 10^{15}\,\, $eV, and
   indirect techniques, the observation of the particle-cascades in the
 atmosphere (extensive air showers: EAS), have to be invoked. The
 information about nature and energy of the primary particles is
 reflected by the shower development \cite{rebel2}
 whose details and signatures for the
 primary particle depend on the high-energy nuclear interactions,
governing the cascading processes. Thus the analysis requires a
 reliable description of these processes, formulated as a hadronic
 interaction model which can be used as generator of Monte-Carlo
  simulations of air showers. It should describe the currently available
 experimental information from accelerator experiments (in particular the
 data from the large collider facilities at CERN and Fermilab) and allow
 a justified extrapolation to experimentally unexplored energy regions.
In the case of the EAS cascades, the quest is for the cross sections
(multiparticle production, rapidity and transverse momentum
 distributuions) for hadron-hadron, hadron-nucleus and nucleus-nucleus
collisions as function of the energy (from pion production threshold up
to ultrahigh energies), most importantly for the forward fragmentation
 region, while actually the central region of the collisions is best
 studied at accelerators. The fragmentation region is nevertheless also
 relevant for the interaction models describing the experimental
 observations at SPS energies and beyond
\cite{topor2},\cite{ranf94a},\cite{ranf94b}.

 There are many hadronic transport models en vogue which address this
 problem. They comprise versions of the Dual-Parton model (DPM)
 \cite{dpm94},
 Quark-Gluon String models (QGSM) \cite{ame1},
 and models designated with the
 name of the code like VENUS  \cite{wer7},
 FRITIOF \cite{ander},
 HIJING \cite{wang0}-\cite{wang3} ,Parton Cascade Models
 \cite{geiger95},\cite{amelin95} and others.
 Some have been specifically developed as Monte-Carlo generator for air
 shower simulations at cosmic ray energies like DPM \cite{capde89},
  HEMAS \cite{forti90} and SYBILL \cite{flet94}.

 Recently \cite{schatz1} the VENUS approach,
 linked to the CORSIKA code \cite{capde92} (now
 widely used for cosmic rays EAS simulations) has been used to
 scrutinize the superposition hypothesis in nucleus-nucleus collisions
 at ultrahigh energies. The hypothesis has been shown to be a rather good
 approximation, for the air shower cascade while Ranft
 \cite{ranf94b} using the
 DPMJET-II version of the DPM approach and considering the fragmentation
 and central regions with equal importance concluded that the
superposition is a rather rough approximation of the reality. Our
 present work is based on the experience with a model and its extensions
 which are the basis of the HIJING code and used for an extrapolation of
 the particle production dynamics from proton-proton (pp) to
 proton-nucleus (pA) and nucleus-nucleus (AA) interactions, taking into
 account the essential constraints of geometry and kinematics. At higher
 collision energies semi-hard processes are included by pertubative QCD
 processes (pQCD).

While the HIJING model ignores final state interactions, the VENUS model
includes reinteractions of string segments among themselves and with
spectator matter, with occurrence of "double strings" which mediate
multinucleon interactions in nucleus-nucleus collisions.

 Both string-models have been applied to a variety of pp, pA and AA
 collision data (see references \cite{wer7},\cite{wang1},\cite{wang2},
 \cite{topor1}.
 However, a consistant
intercomparison of predictions of multiparticle production, transverse
momentum and rapidity distributions at ultrahigh energies and with
respect to their relevance in the EAS cascade is missing. The present
 paper is a first attempt of such a comparison, revealing the most
 salient features and differences in a selected number of cases. We
 introduce the presentation of numerical results with a brief reminder
 of the basis of the HIJING model under consideration, stressing the
 different procedures in defining the interacting nucleon configurations,
 the quark-gluon string formation and the decay into secondary particles.

 The models have been tested at accelerator energies
 for proton-proton and nucleus-nucleus interactions
 and then theoretical predictions on pseudorapidity distributions
 of transverse energy,transverse momenta and secondary particles
 spectra as well as their energy and mass dependence are given
 using HIJING model for proton - Air Nucleus (p+Air) interactions
 between 1 TeV - 1000 TeV and for Nucleus - Air (A+Air) interactions
 at 17.86 TeV/Nucleon corresponding to 1 PeV for iron (Fe) nucleus.
 A comparison with recent results \cite{schatz1}
 at the same energy  using VENUS model is
 also done.
 We have investigated  Feynman scaling behaviour of the
 model in this energy region ,and
the multiple minijets production is considered  for study
  of the charged multiplicity distributions.
  Finally a brief discussion on validity of superposition
  models ,taken into consideration mean integrated values
  of transverse energy predicted by HIJING model,is presented.

\section{Outline of  HIJING   Model }

A detailed discussion of the HIJING Monte Carlo model was reported
in references \cite{wang0}-\cite{wang3}.
The formulation of HIJING was guided by the LUND-FRITIOF and Dual
Parton Model(DPM)  phenomenology for soft nucleus-nucleus reactions at
intermediate energies ($\sqrt{s}<20\,\, GeV$) and implementation
pQCD processes in the PHYTHIA model\cite{sjos94}
 for hadronic interactions.
We give in this section a brief review of the aspect of the model
relevant to hadronic interaction:

\begin{enumerate}
\item Exact diffuse nuclear geometry is used to calculate the impact
parameter dependence of the number of inelastic processes.
\item Soft beam jets are modeled by quark-diquark strings with gluon
kinks along the lines of the DPM and FRITIOF models. Multiple low
$\,p_{T}\,$ exchanges among the end point constituents are included.
\item  The model includes multiple mini-jet production with initial
and final state radiation along the lines of the PYTHIA model
 and with cross sections calculated
within the eikonal formalism.
\item The hadronization of single chains is handled by the LUND code
 JETSET 7.3 \cite{sjos94},that summarizes data on $e^+e^-$.
 In this picture , each nucleon-nucleon collision results in
 excitation of the nucleon by the stretching of a string between the
 valence quark and diquark ( longitudinal excitation ,
  without colour exchange).A phenomenological excitation function
  determines the mass and momentum of the string after each
  interaction.After the last interaction the string decays to
  produce particles.
\item HIJING does not incorporate any  mechanism for final state
interactions among low $\,p_{T}\,$ produced particles
nor does it have colour rope formation.
\end{enumerate}

Unlike heavy ion collisions
at the existing AGS/BNL and SPS/CERN energies, most of the physical
processes occurring at very early times in the violent collisions of
heavy nuclei at cosmic ray energies
 involve hard or semihard parton scatterings
which will result in enormous amount of jet production and
can be described in terms of pQCD.
Assuming  independent production, it has been shown that
the multiple minijets production is important in proton-antiproton
($p\bar{p}$) interactions to account for the increase
of total cross section \cite{gaisser} and the violation
of Koba-Nielsen-Olesen (KNO) scaling of the charged
multiplicity distributions \cite{wang1}.

        In high energy heavy ion collisions, minijets
have been estimated \cite{kaja} to produce 50\% (80\%) of
the transverse energy in central heavy ion collisions at
RHIC (LHC) energies. While not resolvable as distinct jets,
they would lead to a wide variety of correlations, as in
proton-proton($pp$) or antiproton-proton $\bar{p}p$ collisions,
among observables such as
multiplicity, transverse momentum, strangeness, and
fluctuations that compete with the expected signatures
of a QGP. Therefore, it is especially important to
calculate these background processes.

       These processes are calculated in pQCD starting with
 the cross section of hard parton scatterings \cite{eichten} :
\begin{equation}
        \frac{d\sigma_{jet}}{dP_T^2dy_1dy_2} =
        K\sum_{a,b} x_1 x_2 f_a(x_1,P_T^2)f_b(x_2,P_T^2)
        d\sigma^{ab}(\hat{s},\hat{t},\hat{u})/d\hat{t}, \label{eq:sjet1}
\end{equation}
where the summation runs over all parton species, $y_1$,$y_2$
are the rapidities of the scattered partons and $x_1$,$x_2$ are
the fractions of momentum carried by the initial partons and
they are related by $x_1=x_T(e^{y_1}+e^{y_2})/2$,
$x_2=x_T(e^{-y_1}+e^{-y_2})$, $x_T=2P_T/\sqrt{s}$. A factor,
$K\approx 2$ accounts roughly for the higher order corrections.
The default structure functions, $f_a(x,Q^2)$, in HIJING are taken
to be Duke-Owens structure function set 1 \cite{duke}.

        Integrating Eq.~\ref{eq:sjet1} with a low $P_T$
cutoff $P_0$, one abtain  the total inclusive jet cross
section $\sigma_{jet}$. The average number of semihard parton collisions
for a nucleon-nucleon collision at impact parameter $b$ is
$\sigma_{jet}T_N(b)$, where $T_N(b)$ is partonic overlap function
between the two nucleons. In terms of a semiclassical
probabilistic model \cite{gaisser,wang1}, the probability
for multiple minijets production is :
\begin{equation}
        g_j(b)=\frac{[\sigma_{jet}T_N(b)]^j}{j!}e^{-\sigma_{jet}T_N(b)},\;\;
                j\geq 1. \label{eq:sjet3}
\end{equation}
Similarly,  the soft interactions  are represented by
an inclusive cross section $\sigma_{soft}$ which, unlike
$\sigma_{jet}$, can only be determined phenomenologically.
The probability for only soft interactions without any hard
processes is :
\begin{equation}
        g_0(b)=[1-e^{-\sigma_{soft}T_N(b)}]e^{-\sigma_{jet}T_N(b)}.
                \label{eq:sjet4}
\end{equation}
 and the total inelastic cross section for nucleon-nucleon
collisions :
\begin{eqnarray}
        \sigma_{in}&=&\int{d^2b}\sum_{j=0}^{\infty}g_j(b) \nonumber \\
        &=&\int{d^2b}[1-e^{-(\sigma_{soft}+\sigma_{jet})T_N(b)}].
                \label{eq:cin}
\end{eqnarray}
Define a real eikonal function,
\begin{equation}
        \chi(b,s)\equiv\frac{1}{2}\sigma_{soft}(s)T_N(b,s)+
                        \frac{1}{2}\sigma_{jet}(s)T_N(b,s), \label{eq:eiko}
\end{equation}
 the elastic, inelastic, and total cross sections of
nucleon-nucleon collisions are given by :
\begin{equation}
        \sigma_{el}=\pi\int_{0}^{\infty}db^2\left[1-
                e^{-\chi(b,s)}\right]^2, \label{eq:cin1}
\end{equation}
\begin{equation}
        \sigma_{in}=\pi\int_{0}^{\infty}db^2\left[1-
                e^{-2\chi(b,s)}\right],\label{eq:cin2}
\end{equation}
\begin{equation}
        \sigma_{tot}=2\pi\int_{0}^{\infty}db^2\left[1-
                e^{-\chi(b,s)}\right],\label{eq:cin3}
\end{equation}
 Assuming that the parton density in a nucleon can be
approximated  by the Fourier transform of a dipole form factor,
the overlap function is :
\begin{equation}
        T_N(b,s)=2\frac{\chi_0(\xi)}{\sigma_{soft}(s)},\label{eq:over1}
\end{equation}
with
\begin{equation}
        \chi_0(\xi)=\frac{\mu_0^2}{96}(\mu_0 \xi)^3 K_3(\mu_0 \xi),
                \;\; \xi=b/b_0(s),\label{eq:over2}
\end{equation}
where $\mu_0=3.9$ and $\pi b_0^2(s)\equiv\sigma_0=\sigma_{soft}(s)/2$
is a measure of the geometrical size of the nucleon. The
eikonal function  can be written as :
\begin{equation}
        \chi(b,s)\equiv\chi(\xi,s)
        =[1+\sigma_{jet}(s)/\sigma_{soft}(s)]\chi_0(\xi).
\end{equation}

        $P_0\simeq 2$ GeV/$c$ and a constant value of
$\sigma_{soft}(s)=57$ mb are chosen to fit the experimental
data on cross sections \cite{wang1} in $pp$ and $p\bar{p}$
collisions. The equations listed above are used to simulate
multiple jets production at the level of nucleon-nucleon
collisions in HIJING Monte Carlo program. Once the number
of hard scatterings is determined, PYTHIA  algorithms
 generate the kinetic variables of the scattered partons
 and the initial and final state radiations.

 After all  binary collisions are processed,
 the scattered partons in the associated nucleons
 are connected with the corresponding
valence quarks and diquarks to form string systems. The strings
are then fragmented into particles.

   The HIJING model incorporate  nuclear effects such
   as parton shadowing and jet quenching.HIJING is designed
   also to explore the range of possible initial conditions
   that may occur in nuclear collisions at Cosmic Ray
    and colliders  (RHIC,LHC) energies.

   To include the nuclear effects on jet production and
fragmentation,  the EMC \cite{shadow1}
effect of the parton structure functions in nuclei and the
interaction of the produced jets with the excited nuclear
matter in heavy ion collisions are considered.

 It has been observed \cite{shadow1}
that the effective number of quarks and antiquarks in a
nucleus is depleted in the low region of $x$.
While theoretically there may be differences between
quark and gluon shadowing,one assume that the shadowing
effects for gluons and quarks are the same.

        At this stage, the experimental data unfortunately
can not fully determine the $A$ dependence of the shadowing.
 In the HIJING model the A dependence are taken from Ref. \cite{shadow2}:
\begin{eqnarray}
        R_A(x)&\equiv&\frac{f_{a/A}(x)}{Af_{a/N}(x)} \nonumber\\
         &=&1+1.19\ln^{1/6}\!A\,[x^3-1.5(x_0+x_L)x^2+3x_0x_Lx]\nonumber\\
           & &-[\alpha_A-\frac{1.08(A^{1/3}-1)}{\ln(A+1)}\sqrt{x}]
                        e^{-x^2/x_0^2},\label{eq:shadow}\\
        \alpha_A&=&0.1(A^{1/3}-1),\label{eq:shadow1}
\end{eqnarray}
where $x_0=0.1$ and $x_L=0.7$. The term proportional to $\alpha_A$ in
Eq.~\ref{eq:shadow} determines the shadowing for $x<x_0$ with the
most important nuclear dependence, while the rest gives the overall
nuclear effect on the structure function in $x>x_0$ with some very slow
$A$ dependence.

        To take into account  impact parameter dependence,
 one assume that the shadowing effect $\alpha_A$ is proportional
to the longitudinal dimension of the nucleus along the straight
trajectory of the interacting nucleons. The values of
$\alpha_A$ in Eq.~\ref{eq:shadow} are parametrized like :
\begin{equation}
        \alpha_A(r)=0.1(A^{1/3}-1)\frac{4}{3}\sqrt{1-r^2/R_A^2},
                        \label{eq:rshadow}
\end{equation}
where $r$ is the transverse distance of the interacting
nucleon from its nucleus center and $R_A$ is the radius of the
nucleus. For a sharp sphere nucleus with overlap function
$T_A(r)=(3A/2\pi R_A^2)\sqrt{1-r^2/R_A^2}$, the averaged
$\alpha_A(r)$ is $\alpha_A=\pi\int_0^{R_A^2}dr^2 T_A(r)\alpha_A(r)/A$.

        To simplify the calculation during the Monte Carlo
simulation, one can decompose $R_A(x,r)$ into two parts,
\begin{equation}
        R_A(x,r)\equiv R_A^0(x)-\alpha_A(r)R_A^s(x),
\end{equation}
where $\alpha_A(r)R_A^s(x)$ is the term proportional to $\alpha_A(r)$
in Eq.~\ref{eq:shadow} with  $\alpha_A(r)$ given in Eq.~\ref{eq:rshadow}
and $R_A^0(x)$ is the rest of $R_A(x,r)$. Both $R_A^0(x)$ and $R_A^s(x)$
are now independent of $r$.

The effective jet production cross section
of a binary nucleon-nucleon  interaction in $A+B$ nuclear collisions
is then,
\begin{equation}
    \sigma_{jet}^{eff}(r_A,r_B)=\sigma_{jet}^0-\alpha_A(r_A)\sigma_{jet}^A
          -\alpha_B(r_B)\sigma_{jet}^B
      +\alpha_A(r_A)\alpha_B(r_B)\sigma_{jet}^{AB},\label{eq:sjetab}
\end{equation}
where $\sigma_{jet}^0$, $\sigma_{jet}^A$, $\sigma_{jet}^B$ and
$\sigma_{jet}^{AB}$ can be calculated through Eq.~\ref{eq:sjet1} by
multiplying \\
$f_a(x_1,P_T^2)f_b(x_2,P_T^2)$ in the integrand with
$R_A^0(x_1)R_B^0(x_2)$, $R_A^s(x_1)R_B^0(x_2)$, $R_A^0(x_1)R_B^s(x_2)$ and
$R_A^s(x_1)R_B^s(x_2)$ respectively.
With calculated values of
$\sigma_{jet}^0$, $\sigma_{jet}^A$, $\sigma_{jet}^B$ and $\sigma_{jet}^{AB}$,
one get the effective jet cross section
$\sigma_{jet}^{eff}$ for any binary nucleon-nucleon collision.

        Another important nuclear effect on the jet
production in heavy ion collisions is the final state
integration. In  high energy heavy ion collisions, a dense
hadronic or partonic matter must be produced in the central
region. Because this matter can extend over a transverse
dimension of at least $R_A$, jets with large $P_T$ from
hard scatterings have to traverse this hot environment. For
the purpose of studying the property of the dense matter
created during the nucleus-nucleus collisions,
HIJING model include an option to model jet quenching in
terms of a simple gluon splitting mechanism \cite{wang1},\cite{wang2}.

 The induced radiation in HIJING  model is given via a simple
collinear gluon splitting scheme with  energy loss $dE/dz$.
The energy loss for gluon jets is twice that of quark jets\cite{dedx}.
 One assume that interaction only occur with the locally comoving
matter in the transverse direction. The interaction points are
determined via a probability :
\begin{equation}
        dP=\frac{d\ell}{\lambda_s}e^{-\ell/\lambda_s},
\end{equation}
with given mean free path $\lambda_s$, where $\ell$ is the
distance the jet has traveled after its last interaction.
The induced radiation is simulated by transferring a part of
the jet energy $\Delta E(\ell)=\ell dE/z$ as a gluon kink to
the other string which the jet interacts with.
The procedure  is repeated until the jet is out
of the whole excited system
or when the jet energy is smaller than a cutoff below which
a jet can not loss energy any more. This cutoff is taken as
the same as the cutoff $P_0$ for jet production. To determine
how many and which excited strings could interact with the
jet, one  assume a cross section of jet interaction so that
excited strings within a cylinder of radius $r_s$ along the jet
direction could interact with the jet. $\lambda_s$
should be related to $r_s$ via the density of the system of excited
strings. The values for $\lambda_s$ and $r_s$ are taken as
two parameters in the model.

  The main usefulness of these schematic approaches
 for nuclear shadowing and jet quenching is to test
the sensitivity of the final particle spectra .

 For a detailed description of the VENUS model as applied
in the calculations presented in this paper we refer to
excelent review of Werner (sections 6-10)  \cite{wer7}.

\section{NUMERICAL RESULTS}

 For  used a Monte Carlo generator in Cosmic Ray
 physics this model should provide the basic hadronic interaction
 term for the Cosmic Ray cascade i.e should provide the cross
 sections for hadron-hadron,hadron-nucleus and nucleus-nucleus
 collisions as function of the energy .
 Also should provide
 a good description of secondary particles production since,
 secondary $\pi^{0}$ and $\eta$ mesons are the source of the
 electromagnetic shower , secondary $\pi^{\pm}$ and $K^{\pm}$
  mesons are the source of Cosmic Ray Muons and the source of
  atmospheric Neutrinos produced by Cosmic Ray cascade  and
  secondary charmed mesons are the source for prompt Muons
  and Neutrinos. The model should work  from accelerator
  energies up to the highest possible primary energies .
  Experimental observations like rapidity plateaus and
  average transverse momenta rising with energy,KNO scaling
  violation,transverse momentum - multiplicity correlations
  and minijets pointed out that soft hard and processes are closely
  related and all these properties were understood within the
  HIJING model \cite{wang0}-\cite{wang3}.

  The HIJING model provides a framework not only for the study
  of hadron-hadron interactions ,but also for the description of
  particle production in hadron nucleus and nucleus - nucleus
  collisions at ultrahigh energies.

  The relevance of an event generator like HIJING or VENUS
  for hadron production cross sections in the Cosmic
  Ray energy region can only be claimed if the model
  agrees to the best available data in the accelerator energy
  range and if it shows a smooth behaviour in the extrapolation
  to ultrahigher energies .

  For the Cosmic Ray Cascade in the atmosphere only hadron -
   nucleus  and nucleus-nucleus collisions are relevant
   with Nitrogen ${^{14}}N$ beeing the most
   important target nucleus.
   However experimental data are of much better quality in
   hadron-hadron,and especially in proton-proton collisions
   or antiproton-proton collisions than for collisions
   of hadrons with  light nuclei.The scarcity of the data
   in this region was also pointed out in studing
   strangeness production at SPS energies \cite{topor1}.

   So we start with the study of proton - proton collisions and
   nucleus - nucleus collisions at accelerator energies .

\subsection{ PROTON - PROTON AND NUCLEUS-NUCLEUS
INTERACTIONS  AT ACCELERATOR ENERGIES}

We used  the program HIJING with default parameters:
\begin{enumerate}
\item IHPR2(11)=1 gives the baryon production
model with diquark-antidiquark pair production allowed, initial
diquark treated as unit;
\item IHPR2(12)=1, decay of particle such as
$\,\pi^{0}\,$,$\,K_{s}^{0}\,$,$\,\Lambda\,$,$\,\Sigma\,$,
$\,\Xi\,$, $\,\Omega\,$ are allowed\,\,;\,\,
\item IHPR2(17)=1 - Gaussian distribution of transverse
momentum of the sea quarks ;
\item IHPR2(8)=0 - jet production turned
off for theoretical predictions denoted by HIJING model ;
\item IHPR2(8)=10-the maximum
number of jet production per nucleon-nucleon interaction
for  for theoretical predictions denoted by
$\,HIJING^{(j)}\,$  at SPS energies and 300 GeV ;
\item IHPR2(8)=20  for energies $\geq 1\,\, TeV $ .
 We have neglected  also
 nuclear shadowing effect,and jet quenching
(see section 2) and  we have a cut in pseudorapidity
 $\eta >  0.5 $ for ultrahigh energies ($\geq 1\,\, TeV $).
 \end{enumerate}

In Table I the calculated average multiplicities of particle at
$E_{lab}=200\,\, GeV\,$ in ($pp$) interactions are compared to data.
The theoretical values as predicted by the model
$\,HIJING\,$ and  $\,HIJING^{(j)}\,$
are obtained for
$\,\,10^{5}\,\,$ generated events and in a full phase space.
The values $\,\,HIJING^{(j)}\,\,$ include the
very small possibility of mini jet production at these
low SPS energies.
The experimental data are taken from Gazdzicki and Hansen
\cite{1na35}.
 The theoretical values given by  VENUS model(as computed here)
 are obtained for
 $\,\,10^{4}\,\,$ generated events and in a full phase space.
 The values labeled DPMJET II are taken from
Ranft \cite{ranf94a}.

The small kaon to pion ratio is due to the suppressed
strangeness production basic to string fragmentation.
Positive pions and
kaons are  more abundant than the negative ones due to
charge conservation. We note that the {\em integrated}
multiplicities for neutral
strange particle $\,\,<\Lambda>,<\bar{\Lambda}>,<K_{s}^{0}>\,\,$
are reproduced at the level of three standard  deviations for
 $\,pp\,$ interactions at  $\,200\, GeV\,$ and 300 GeV.
 (see also Fig.1a and Fig.1b -experimental data from Lo Pinto et al.
  \cite{ex80} for $\,pp\,$ interactions at 300 GeV.)
 However the values for
$\,<\bar{p}>\,$  and $\,<\bar{\Lambda}>\,$ are significantly
 over predicted  by the models.

For completeness
we include a comparison of hadron yields
at collider energies $\,E_{LAB}=160\,\, TeV \, $
for $\,\,\bar{p}p\,\,$ interactions, where mini-jet
production plays a much more important role.
{}From different collider experiments
Alner et al. (UA5 Collaboration) \cite{1ua5}  attempted to piece
 together a picture of the composition of a typical soft event at the
 Fermilab $\,Sp\bar{p}S\,$  collider \cite{ward}.
The measurements were made in various
different kinematic regions and have been extrapolated in the full
transverse momenta($\,p_{T}\,$)  and rapidity range for comparison
as described in reference \cite{1ua5}.
The experimental data are compared to theoretical values obtained with
$HIJING^{(j)}$ in Table II.
 It was stressed  by  Ward \cite{ward} that
 the data show a substantial excess of photons compared to the
 mean  value for pions $\,\,<\pi^{+}+\pi^{-}>\,\,$.
 It was suggested as a possible
 explanation of such enhancement a  gluon
 Cerenkov radiation emission in hadronic collision
\cite{drem}. Our calculations rules out such hypothesis.
 Taking into account decay from resonances and
 direct gamma production,  good
agreement is found within the experimental errors.

In the following plots
the kinematic  variable used to  describe
single particle properties are
 the transverse momentum $\,p_{T}\,$ and the rapidity $\,y\,$ defined
as usual as:
 \vskip 0.3cm
 \begin{equation}
 y=\frac{1}{2}ln \frac{E+p_{3}}{E-p_{3}}=ln\frac{E+p_{3}}{m_{T}}
 \label{e18}
 \end{equation}
 \vskip 0.3cm
 with $\,E,p_{3}\,$,and $\,m_{T}\,$ being energy,longitudinal momentum and
 transverse mass
$ m_{T}=\sqrt{m_{0}^{2}+p_{T}^{2}}$
 with $\,m_{0}\,$ being the particle rest mass.

  The pseudorapidity $ \eta $ is used rather than the rapidity
  since for $ \eta $  no knowledge of particle masses is
  required.
 \vskip 0.3cm

 \begin{equation}
 \eta = \frac{1}{2}ln \frac{p+p_{3}}{p-p_{3}}= -
 ln tan{\frac{\theta}{2}}
 \label{e19}
 \end{equation}
 \vskip 0.3cm
  where $p$ is  the projectile nucleon momentum and $\theta$ is the
  scattering angle.

 Feynman $x_F$ variable is defined for ultrahigh energy as:
 \vskip 0.3cm
 \begin{equation}
 x_F=2 \frac{m_T}{\sqrt{s}}sinh(y^{cm})
 \label{e20}
 \end{equation}
 \vskip 0.3cm
  where  $y^{cm}$  ans $s$ are  rapidity
  and total energy in center of mass frame (cms).

 In Figure 1 we compare rapidity and transverse momentum
       distributions for strange particles in proton-proton
      interactions at 300 GeV  given by HIJING model with
   experimental data \cite{ex80}. The agreement is quite good.
   However ,it will  be interesting to investigate in the future
   the Feynman scaling behaviour of the model at the
   accelerator energies since the forward fragmentation
   region seems to play an important role \cite{ranf94a},
   \cite{ranf94b},\cite{topor2}.

  This year marked a real progress in the field of
  high energy nuclear collisions .Now an
  entirely new domain of energies has become accessible
  with heavy nuclear beams \cite{mik95}.
  Recently \cite{topor1} using HIJING and VENUS models
  the systematics of strangeness enhancement was calculated
  for $pp$,$pA$ and $AA$ interactions at SPS energies
  and it was stressed out that the enhancement of strangeness
  has its origins in non-equilibrium dynamics of few
  nucleons systems.To clarify the new physics much better
  quality data on elementary $pp$ as well as other light ion
    reactions will be needed for tunned models .

    We foccused our analyses concerning particle production
    in nucleus - nucleus interactions  at SPS energies ,
    where the models should be better tested ,mainly for
    two simmetrical interactions $S+S$ , $Pb+Pb$ and for
    asymmetrical one $S+Pb$.

    It was shown  that the negative pion rapidity densities
    are well accounted for both HIJING and VENUS \cite{topor2},
    but the flat valence proton distribution in $S+S$ is only
    reproduced by VENUS.Recall that VENUS includes a model
    of final state interactions in dense matter as well as
    a colour rope effect ,called double strings.VENUS
    predicts a much higher degree of baryon stopping at
    midrapidity than HIJING .
    These conclusions are confirmed also for our calculations
    in $S+S$ at $200\,\, AGeV$  (see Figs. 2a,2b) and $Pb+Pb$
    collisions at $160 \,\,AGeV$ (see Figs. 2c,2d) for
    $\,\Lambda\,$ and $\,\bar{\Lambda}\,$ . The experimental
    data are taken from Alber et al. \cite{alber94}.
    The rapidity distributions for antiproton  $\,\bar{p}\,$
    and for negative kaons $K^{-}$ are represented for
    $S+S$ interactions (Fig.3a($\,\bar{p}\,$) and Fig.3b($K^{-}$))
    and $Pb+Pb$ interactions (Fig.3c($\,\bar{p}\,$) and
    Fig.3d ($K^{-}$)) in comparison with some experimental data
    taken from Baechler et al.  \cite{baec93} for $K^{-}$
     and  from Murray et al. \cite{murray95} for $\,\bar{p}\,$ .
      For asymmetrical interactions $S+Pb$ we predict in both models
      the rapidity spectra for $\pi^{-}$ (Fig. 4a) , $K^{-}$
      (Fig. 4b) ,$\,\Lambda\,$ (Fig. 4c) and $\,\bar{\Lambda}\,$
      (Fig. 4d )) at $200\,\, AGeV$ .
     We mention that we have used  VENUS model with the option
     without  decaying  resonances (dotted histograms).
     If will allow decay of
     resonances then  we  get an increase of the values
     especially at mid rapidity which  improve agreement between
     theory and experimental data (dashed histograms).

    It will be interesting to compare these results with upcoming
    data to test if the strangeness enhancement increases from
    $SS$ to $SPb$ or from $SS$ to $PbPb$ interactions.

\subsection{PROTON -AIR NUCLEUS INTERACTIONS
 AT ULTRAHIGH ENERGIES}

 Since not enough data are avaiable in the fragmentation region
 of hadron collisions with light target nuclei,many features of
 particles production in collisions involving nuclei can only be
 extracted from the study of models.
 These kind of analysis have been done recently in VENUS model
 \cite{schatz1} and DPM model \cite{ranf94a},\cite{ranf94b}.
 In HIJING model some specific interactions were investigated
  at RICH and LHC energies \cite{wang1}-\cite{wang3}.

  Taking into account the traditional concept of collisions
  between nucleons and air nuclei the primary loses its energy
  during the interaction and this fluctuate between zero and \\
  100 $\%$ . The fraction of the energy lost and used for producing
  new particles is then referred to as inelasticity.
  We do not investigate traditionaly inelasticities (baryonic,
  electromagnetic or mesonic) see \cite{schatz1}.

  We generate for proton-Air Nucleus ($p+Air$)
  interactions $10^4$ minimum bias events ($b_{min}=0,b_{max}=5 fm$).
  In order to give an idea about energy used for producing
  new particle we investigate in Figure 5 the transverse
  energy pseudorapidities spectra and their
  dependence with energy for all secondaries (Fig. 5a),
  all neutral (Figure 5b) ,all charged (Fig. 5c) and
  gluons (Fig. 5d). As we see from Figure 5  gluons
  carried an important fraction from transverse energy
  and this fraction increase with increasing energy
  (from  2.8 \% at 1 TeV to  17.3 \% at 1000 TeV).
   Also the percentage of occurence of gluons increase
   from 7.20 \% at 17.86 TeV to 10.71 \% at 1000 TeV
   (see table 3).

     The rapidity distributions and energy dependence
     of main secondary particles are shown
     in Figure 6 and Figure 7.Secondary $\pi^{\pm}$ (Fig.6a and 6b)
     and $K^{\pm}$ (Fig.6c and 6d) are the source of Cosmic Ray Muons
     and the source of atmospheric Neutrinos produced by the
     Cosmic Ray cascade. Secondary $\pi^{0}$(Fig.7a) and $\eta$ mesons
     are the source of the electromagnetic shower and secondary
     charmed mesons (Fig. 7d) are the source for prompt Muons
     and Neutrinos . The lambdas have been shown (Fig. 7c),because
     they can be produced from protons by exchanging a single valence
     quark by strange quark.The outer maximum of their distributions
     are due to this process.This component is identified in all
     nucleon distributions (neutron,proton).
     The statistics ($ 10^{4}$ events generated) seems not to be
     enough for charmed mesons ($D^{\pm}$) but we give such
     distributions only to show that introducing HIJING code in
     a shower code with a much higher statistics at simulation
     level a study of promt muon component should be feasible.

     We see from Figure 6 that mesons distributions exhibit a
     broad structureless shape which does not depend strongly
     on the type of mesons but have a dependence on energy
     from  \\
     1 TeV to 1000 TeV which is more pronunciated compared
     with VENUS results (see reference \cite{schatz1}.)

     More less  energy dependence is seen in tranverse momenta
     distributions for secondaries in p+Air Nucleus interactions
     in Figure 8( for all charged particles - Fig. 8a,
      for proton - Fig. 8b, for positive pions $\pi^+$ - Fig. 8c
       and for negative pions $\pi^-$ -Fig.8d)

     Trying to stress out  the relevance of accelerator data
      on particle production in hadron - nucleus collisions for
      Cosmic Ray cascade we study the Feynman scaling behaviour of
      $p+Air \rightarrow \pi^{\pm} + X $  and
      $p+Air \rightarrow p+\bar{p} + X $ in Fig.9a and Fig.9b
      respectively.
      We plot the $x_{F}dN/dx_{F}$ distributions for laboratory
      energies of 1 TeV (dotted histograms),100 TeV (dashed
      histograms),1000 TeV (solid histograms).
      The violations of Feynman scaling which occur
      are connected with
      known rise of rapidity plateau for all kinds of produced
      particles and with production of minijets.
       Due to minijets Feynman scaling is more strongly
       violeted especialy in the region $x_{F} \geq 0$ .
       The violation of Feynman scaling are less dramatic in
       DPMJET II model \cite{ranf94a},\cite{ranf94b} and appear only
       arround $x_{F}=0$ and $x_{F}=1$.We note also that HIJING
      show violation of KNO scaling due to the production of
      multiple minijets and the tendency becomes stronger with
       increasing energy \cite{wang1}.

      In order to evaluate multiplicity distributions for  the
      charged particles in $p+Air$ interactions (Fig.10a)
      we differentiate  the contributions from soft
      (Fig.10b events with $N_{jet}=0$) and hard processes
      (Fig.10c events with $N_{jet}=1$ and
      Fig.10d events with $N_{jet}> 1$ ) where $N_{jet}$ is
      the number of minijets produced in that events.
       Our calculations including the effects of multiple
       minijets  are the contributions from the events which
       have hard colisions with $P_{T} > P_{0}$. Analysing
       Figure 10 it is clear that the events at the tails of
       the charged multiplicity distributions
       in p+Air interactions are mainly those with
       multiple minijets production.

 \subsection{NUCLEUS-AIR NUCLEUS INTERACTIONS AT
    ULTRA-HIGH ENERGIES}

  Due to the large fraction of nuclei in primary Cosmic Rays ,
  Nucleus-Air collisions are of great importance in the
  EAS development.It is important ,that the model will be
  able to give a good description of hadron production in
  nucleus-nucleus interactions.So, in this subsection we
  try to investigate mainly the dependence on projectile mass
  for specific interactions for EAS (He,Ne,S,Fe+Air Nucleus)
  at 17.86 TeV/Nucleon which correspond to 1 PeV laboratory
  energy for Fe nucleus .

  For a real comparison ,the nucleus-nucleus collisions geometry
  should be the same for all of the Monte Carlo codes since
  nuclear density distributions are well known from nuclear
  physics \cite{awes89}.At SPS energies the calculated number
  of target and projectile participants as well as the number
  of participants as a function of reaction impact parameter
   (b fm) shows  no difference in HIJING and VENUS model \cite{topor2}.

 The results depicted in Figure 11 are obtained
 for $10^4$ generated events and for the following
 intervals ($b_{min}-b_{max}$)of impact parameter :
  Fe+Air (0-13 fm), S+Air(0-11 fm),Ne+Air(0-10 fm),
  He+Air (0-7 fm) - dotted histograms ;
  Fe+Air (0- 8 fm) ,S+Air(0-7 fm),Ne+Air(0- 6 fm),He+Air (0-5 fm )
  -dashed histograms ;
  Fe+Air (0 - 5 fm),S+Air(0 - 5 fm),Ne+Air(0 - 5 fm),He+Air (0-4 fm)
   -solid histograms.

  From Figure 11(a,b,c,d) we see that at 17.86 TeV/Nucleon
  the results have strongly dependence  on impact parameter intervals
  ($b_{min}$-$b_{max}$)
  and also the possibility of mixing interactions if we
  are interested in rapidity spectra only.

  Therefore for comparison of HIJING and VENUS results at
   17.86 TeV/Nucleon for p+Air,He+Air,
  Ne+Air,S+Air,Fe+Air and  for ~p+Air at
  ~1000 TeV we try to get approximatively the same number
  of participants. For HIJING model we get the values
  for mean numbers of participants listed in last lines
  of Table 3 for $10^4$ generated events and
   the following impact parameter intervals ($b_{min}-b_{max}$):
  He+Air(0-4 fm),Ne+Air(0-6 fm),S+Air (0-7 fm),Fe+Air (0-8 fm).
   $p+Air$ (0- 5 fm).
  The results for VENUS model are taken from
  Schatz et al. \cite{schatz1}.

  Table 3 lists the frequency of various particles among the
  secondaries for collisions of different nuclei with
  nitrogen. Our calculations confirm the results from
  reference \cite{schatz1} .The percentages do not seem to
  depend strongly on projectile mass nor,as shown by proton
  results on energy. We see only a slight tendency of
  increasing strangeness production with increasing primary
  mass for VENUS results.
  Analysing the results of table 3 we see that the VENUS model
  predict  more pions and kaons , but less gamma particles that
  HIJING model . We remark also differences in total multiplicities
  at ultrahigh energy which can not be explained only by
   the difference between total number of participants .
   It seems that {\em "double string" mechanism change considerably
   the baryon spectra and allows the baryon number to migrate
   several units of rapidity from the end point rapidity}.

   Important for Cosmic Ray studies and EAS development is
   also the dependence of particle production on the nuclear
   target and projectile. Taken into consideration the same
   conditions as those reported for the values listed in
   Table 3 the theoretical predictions for
  mass dependence of pseudorapidities spectra
   are given  for transverse energy in Figure 12,
  for transverse momenta in Figure 13 and
  for main secondaries produced new particles in Figure 14.
  The  transverse momenta  distributions of all charged,
  proton,$\pi^+$,$\pi^-$ are represented in Figure 15.
  For Feynman $x_{F}dN/dx_{F} $ distributions
   we give theoretical values
   for specific EAS interactions in Figure 16 only for
   charged pions and for  sum of protons and antiprotons .
    We see a slight mass dependence of Feynman distributions
    and transverse momenta  distributions for $A \geq 20$ .

    Instead of the proper sampling of Nucleus + Air Nucleus
   scattering events , an approximation often applied in
   EAS development is the so called superposition model.
   There are two different possible superposition models:
   a nucleus-nucleus collision A-B with $N_{part}$
   participating nucleons is approximated as the superposition
   of $N_{part}$ simultaneous nucleon-nucleon collisions and
    the second one : a nucleus-nucleus collision  A-B  with
    $N_{proj}$ participating projectile nucleons is
     approximated as the superposition of $N_{proj}$ simultaneous
     nucleon-B collisions. The validity of this principle
     was analysed  in some recent works \cite{schatz1},
     \cite{ranf94a},\cite{ranf94b} with different conclusions.

    We  investigate  in HIJING model the integrated
    mean transverse energy for secondaries
     produced in Nucleus+Air Nucleus interactions
     at 17.86 TeV/Nucleon. So,we  generate $10^4$ events in the same
     impact parameter interval (0-5 fm) for Fe,S,Ne,He,p+Air
      interactions. The values obtained for mean projectile
     participants ($N_{proj}$) and mean number of
      binary collisions  $N_{coll}$(which include nucleon-nucleon($N-N$),
      nucleon-wounded nucleon($N-N_w$),$N_w-N$ and $N_w-N_w$
      collisions) are listed in Table 4.
     The results given in Table 4  are for all secondaries ,
     all charged and all neutrals particles produced in
     interactions. The  values of integrated
     mean transverse energy $<E_T>$
     scale with number of binary collisions in nucleus-nucleus
     interactions at this energy  $N_{coll}$ and scaling proprietes
      are valid for $A \geq 20$ and should not be
      applied to ligth nucleus+Air interactions.
      Since  integrated mean value of transverse energy $<E_T>$
       is a measurable quantity it will
      be interesting to verify this scaling at RHIC energies.

  In order to evaluate multiplicitity distributions for
  charged particles in A+Air interactions we differentiate
  the contribution for all events (soft + hard)
  (Fig.17a for Fe+Air and Fig.17b for S+Air )
   and only hard processes - events with
  $N_{jet} > 1 $ (Fig.17c for Fe+Air and Fig.17d for
  S+Air ). We can see from Figure 17 that the low multiplicity
  events are dominated by those of no jet production while
  high multiplicity events are dominated by those of at least
  one jet production.
   Also it is clear from Figure 17 that the contributions
  from the events which have hard collisions  increase
  with increasing of available energy and of projectile mass.

   At ultrahigh energies   nuclear effects like
  nuclear shadowing of partons
   and jet quenching (see section 2) ,
   should have important contributions \cite{wang1},\cite{wang2}.
   In HIJING model a simple parametrisation of gluon shadowing
   (see eqs 12-15) and a schematic quenching model
   (see eqs 16-17) were introduced to test the sensitivity
  of the final distributions to these aspects of nuclear
  dynamics.
  Trying to estimate the results taken into consideration  and
   neglected  such effects we expect bigest
  differences for more central Fe+Air interactions
  at 17.86 TeV/Nucleon.Therefore we will estimate these effects
   for impact parameter interval($0-5 fm)$).
  In Figure 18 we give theoretical values for
   pseudorapidity distributions for charged pions (Fig.18a and
  Fig.18b) and for charged kaons (Fig.18c and Fig.18d)
  without parton shadowing and jet quenching (solid histograms)
  and with parton  shadowing
  and jet quenching(dashed histograms).
   The quenching mechanism in HIJING is limited  to
  $p_T > 2 \,\,GeV/c $ partons. For pseudorapidity distributions of
  main secondary particles produced in EAS specific interactions
 in this energy region these effects should be neglected.

  However ,these effects are more pronunciated at the early
  stage of collisions ,like we see from  Figure 19 which
   depicted pseudorapidity distributions for gluons (Fig.19a) and
  quark-antiquark pairs(Fig.19b)
 as well as   transverse momenta distributions
  for valence and sea partons (Fig.19c) and valence parton only
  (Fig.19d).
  Of great importance are also energy densities which are
  produced in these interactions ,at the early collisions
  time  but their values are under study now.

\section{Conclusions}

In this paper we  have performed an  analysis  of  particle
production in $\,pp\,$ and $\,AA\,$ collisions at
SPS CERN-energies using the HIJING and VENUS models
and investigate  possible applications of the models
at Cosmic Ray energies  mainly for
 p+Air Nucleus  at 1 TeV ,10 TeV, 1000 TeV
 and Nucleus-Air Nucleus  interactions (Fe,S,Ne,He+Air,p+Air)
 at $ 17.86 $ TeV/Nucleon .

 We stressed out that integrated multiplicities
 are quite well reproduced at the level of three
 standard deviations for $pp$ interaction at 200 GeV and
 300 GeV. However the values for $<\bar{p}>$ and
 $<\bar{\Lambda}>$ are significantly overpredicted by the
 models. A very good agreement is found within experimental
 errors for ultrahigh energies ( 160 TeV) in HIJING approach
 where mini-jet production plays a much more important role.

  For nucleus-nucleus collisions at SPS energies
 the rapidity spectra are well
  accounted for both HIJING and VENUS models for mesons ,but
  VENUS model seems to give a more good description for
   strangeness  production.At SPS energies a final state
   interactions in dense matter as well as a colour rope
   effect predicts much higher degree of baryon stopping
   at midrapidity than HIJING.

  The event generator VENUS \cite{schatz1} and HIJING in the
  present study were tested to simulate ultrahigh energy
  collisions specific for EAS development.
  The transverse energy,transverse momenta and secondary
 particles produced spectra as well as their energy
  and mass dependences  were investigated in detail.

 The contributions from
 the events which have hard collisions are stongly
 evidentiated at the tail of charged particles multiplicity
 distributions and increase with increasing energy and
 increasing mass of projectile.
 Feynman scaling is  strongly violeted in HIJING model
 due to multiple minijets events.
 For Cosmic Ray energy region and specific EAS interactions
 the effects of  parton shadowing and jet quenching  should
 be neglected.

 Possible utilization inside a shower code of these models
 allows to extend the analyses and to release the simple
 superposition model which is not valid at least for
 integrated mean transverse energy of secondaries particle .
 The thoretical values predicted by
  HIJING model suggest a scaling with
  number of  binary nucleon-nucleon collisions.
  It will be interesting to verify this hypothesis
   in the future experiments at RHIC energies.

\section{Acknowledgements}

We are grateful to Klaus Werner for providing us the source code of VENUS.
One of the authors (VTP) would like to  expresses  his
 gratitude to Professor C.Voci and Professor M.Morando
 for kind invitation and
 acknowledge financial support from INFN-Sezione di Padova,
 Italy where part of the presented calculations were performed.
 VTP  is also indebted to Professor T. Ludlam and M.Gyulassy
 for  kind financial help and hospitality
 during his stay  in Columbia University ,New York
 and  to  Dr. Xin Niang Wang for very useful discussions on
 different aspects of HIJING model.
 He acknowledge also financial support from FZK-Karlsruhe,Germany
 where this work was finished.

\vskip 2cm

\begin{center}
{\large \bf Figure Captions}
\end{center}

\begin{figure}[h]
\caption{Rapidity (Fig.1a) and transverse momenta(Fig.1b)
distributions of $\,\Lambda^0\,$(dotted histograms),
$\,\bar{\Lambda^0}\,$(dashed histograms),
and $K_{S}^{0}$ (full histograms)
  produced in $pp$ interactions at $\,300\,\,GeV\,$ .
 HIJING results are shown by histograms.
 The experimental data are taken from \protect{\cite{ex80}}.}
\end{figure}

\begin{figure}[h]
\caption{Rapidity distributions of  $\Lambda$ and
  $\bar{\Lambda}$ produced  in  central
 $SS$ collisions at $200$ AGeV (Fig.2a and Fig.2b) respectively
  and in  central $PbPb$ collisions at $160$ AGeV
 (Fig.2c and Fig.2d respectively).Expectations based on HIJING
 model are depicted as solid  histograms .
 The theoretical predictions based on VENUS model
 are depicted as dotted histograms (option without decaying of
 resonances) and as dashed histograms (option with decay of
 resonances).
  The NA35 data (full circles) are from Alber et al.
  \protect{\cite{alber94}}. The open circles show
  the distributions for $\,SS\,$ collisions
   reflected at $\,y_{lab}=3.0\,$.}
\end{figure}

\begin{figure}
\caption{ Comparison of central $S+S$
  collisions at $\,200\,$ AGeV
 (Fig.3a -antiproton and Fig.3b - negative kaons)
 with central $PbPb$ collisions at $\,160\,$ AGeV
 (Fig.3c-antiproton and Fig.3d- negative kaons).
 The experimental data are from NA44 \protect{\cite{murray95}} for
 antiprotons and from \protect{\cite{baec93}} for negative kaons.
 The open circles show the distributions
 for $\,SS\,$ collisions reflected at $\,y_{lab}=3.0\,$.
 The solid,dashed and dotted histograms have the same
 meaning as in Figure 2.}
\end{figure}
\begin{figure}
\caption{ The various theoretical predictions for
 $S+Pb \rightarrow \pi^- +X$(Fig.4a),
 $S+Pb \rightarrow K^-+X$(Fig.4b),
 $S+Pb \rightarrow \Lambda +X$ (Fig.4c) and
 $S+Pb \rightarrow \bar{\Lambda} +X$ (Fig.4d) at
 $200 \,\,AGeV$. The solid and dotted histograms have the same
 meaning as in Figure 2.}
\end{figure}
\begin{figure}
\caption{ Pseudorapidity distributions for transverse energy of
secondary produced particles: all secondaries (Fig.5a) ,
 all neutrals( Fig-5b) , all charged (Fig.5c)  and gluons
 (Fig.5d) in  $p+Air$ Nucleus interactions .
The dotted(for 1 TeV laboratory energy) ,
dashed(for 100 TeV laboratory energy),
and solid(for 1000 TeV laboratory energy) histograms are theoretical
values given by HIJING  model.}
\end{figure}
\begin{figure}
\caption{Pseudorapidity distributions for
 $p+Air \rightarrow \pi^+ + X $ (Fig.6a),
 $p+Air \rightarrow \pi^-+ X $ (Fig.6b),
 $p+Air \rightarrow K^+ + X $ (Fig.6c),
 $p+Air \rightarrow K^- + X $ (Fig.6d) , at 1 TeV,100 TeV and 1000 TeV.
 The dotted ,dashed and solid histograms have the same meaning as in
 Figure 5.}
\end{figure}
\begin{figure}
\caption{Pseudorapidity distributions for
 $p+Air \rightarrow \pi^o + X $ (Fig.7a),
  $p+Air \rightarrow \gamma+ X $ (Fig.7b),
 $p+Air \rightarrow \Lambda^0 + X $ (Fig.7c),
  $p+Air \rightarrow D^{\pm} + X $ (Fig.7d),at
 1 TeV,100 TeV and 1000 TeV.
   The dotted ,dashed and solid histograms have
   the same meaning as in Figure 5.}
\end{figure}
\begin{figure}
\caption{Transverse momenta distributions for
 $p+Air \rightarrow all\,\, charged + X $ (Fig.8a),
 $p+Air \rightarrow p +X $ (Fig.8b),
 $p+Air \rightarrow \gamma+ X $ (Fig.8c),
 $p+Air \rightarrow \bar{p} + X $ (Fig.8d),
 at 1 TeV,100 TeV and 1000 TeV.
 The dotted ,dashed and solid histograms have
 the same meaning as in Figure 5.}
\end{figure}
\begin{figure}
\caption{Test of Feynman scaling in the production of
 $p+Air\rightarrow \pi^{\pm}+X$ collisions
 (Fig.9a) and $p+Air\rightarrow p+\bar{p} + X$ collisions (Fig.9b),
 between 1 TeV - 1000 TeV.
 The Feynman -$ x_F$ distributions were calculated with HIJING model.
  The dotted ,dashed and solid histograms have
 the same meaning as in Figure 5.}
\end{figure}
\begin{figure}
\caption{Charged multiplicities distributions in $p+Air$
   interactions at 1000 TeV.Contibutions from all events
  are depicted in Fig.10a.
    In Fig.10b the histogram  is  from HIJING model
   calculations with contributions from events with $N_{jet}=0$,
   Fig.10c the histogram is  from calculations with
   contributions from events with $N_{jet}=1$ and in
   Fig.10d  the histogram is from calculations with
   contributions from events with $N_{jet}>1$.}
\end{figure}

\begin{figure}
\caption{Pseudorapidity distributions for all secondaries produced
 in Fe+Air interactions(Fig.11a),S+Air interactions(Fig.11b),
 Ne+Air interactions(Fig.11c)and He+Air interactions (Fig.11d)
 at 17.86 TeV/Nucleon for events generated in
 different impact parameters intervals ($b_{min},b_{max}$).
 See the text for explanations.}
\end{figure}

\begin{figure}
\caption{Pseudorapidity distributions for transverse energy of
 secondary produced particles :all secondaries(Fig.12a);
 all neutrals(Fig.12b);all charged(Fig.12c);gluons(Fig.12d)
 in $A+Air$ interactions at 17.86 TeV/Nucleon.
 The theoretical values were calculated with HIJING model
 and are depicted by dotted (He+Air) ; dot-dashed (Ne+Air) ;
 dashed (S+Air) and solid (Fe+Air) histograms.}
\end{figure}

\begin{figure}
\caption{Pseudorapidity distributions for transverse momenta of
secondary particles for $A+Air\rightarrow \pi^{\pm}+X$ (Fig.13a)
 and $A+Air \rightarrow p+X$ colisions (Fig.13b)
  at 17.86 TeV/Nucleon.
 The histograms have the same meaning as in Figure 12.}
\end{figure}

\begin{figure}
\caption{Pseudorapidity distributions for
  $A+Air \rightarrow \pi^+ + X $ (Fig.14a),
 $A+Air \rightarrow \pi^-+ X $ (Fig.14b),
 $A+Air \rightarrow K^+ + X $ (Fig.14c),
 $A+Air \rightarrow K^- + X $ (Fig.14d), at 17.86 TeV/Nucleon.
 The histograms have the same meaning as in Figure 12.}
\end{figure}

\begin{figure}
\caption{Transverse momenta distributions for
  $A+Air \rightarrow all\,\, charged + X $ (Fig.15a),
  $A+Air \rightarrow p +X $ (Fig.15b),
  $A+Air \rightarrow \pi^++ X $ (Fig.15c),
  $A+Air \rightarrow \pi^- + X $ (Fig.15d),at 17.86 TeV/Nucleon.
 The histograms have the same meaning as in Figure 12.}
\end{figure}

\begin{figure}
\caption{ The Feynman $x_{F}$ distributions in the production of
  $A+Air \rightarrow \pi^{\pm}+X$ collisions (Fig.16a) and
  $A+Air \rightarrow p+\bar{p}+X$ colisions(Fig.16b), at 17.86
  TeV/Nucleon.
  The histograms have the same meaning as in Figure 12 .}
\end{figure}

\begin{figure}
\caption{Charged particles multiplicities distributions in
  Fe+Air interactions (Fig.17a) and S+Air interactions
  (Fig.17b) at 17.86 TeV/Nucleon.
  Contributions from events with number of minijets
   $N_jet > 1$ are represented in Fig.17c for Fe+Air interactions
   and in Fig.17d for S+Air interactions.}
\end{figure}

\begin{figure}
\caption{ Pseudorapidity distributions for
   $Fe+Air \rightarrow \pi^+ + X $ (Fig.18a),
   $Fe+Air \rightarrow \pi^- + X $ (Fig.18b),
   $Fe+Air \rightarrow K^+ + X $ (Fig.18c),
   $Fe+Air \rightarrow K^- + X $ (Fig.18d) at
    17.86 TeV/Nucleon and impact parameter interval (0-5 fm).
   Solid histograms are HIJING model predictions.Also
   showing the influence of nuclear shadowing and
   jet quenching effects (dashed histograms).}
\end{figure}
\newpage

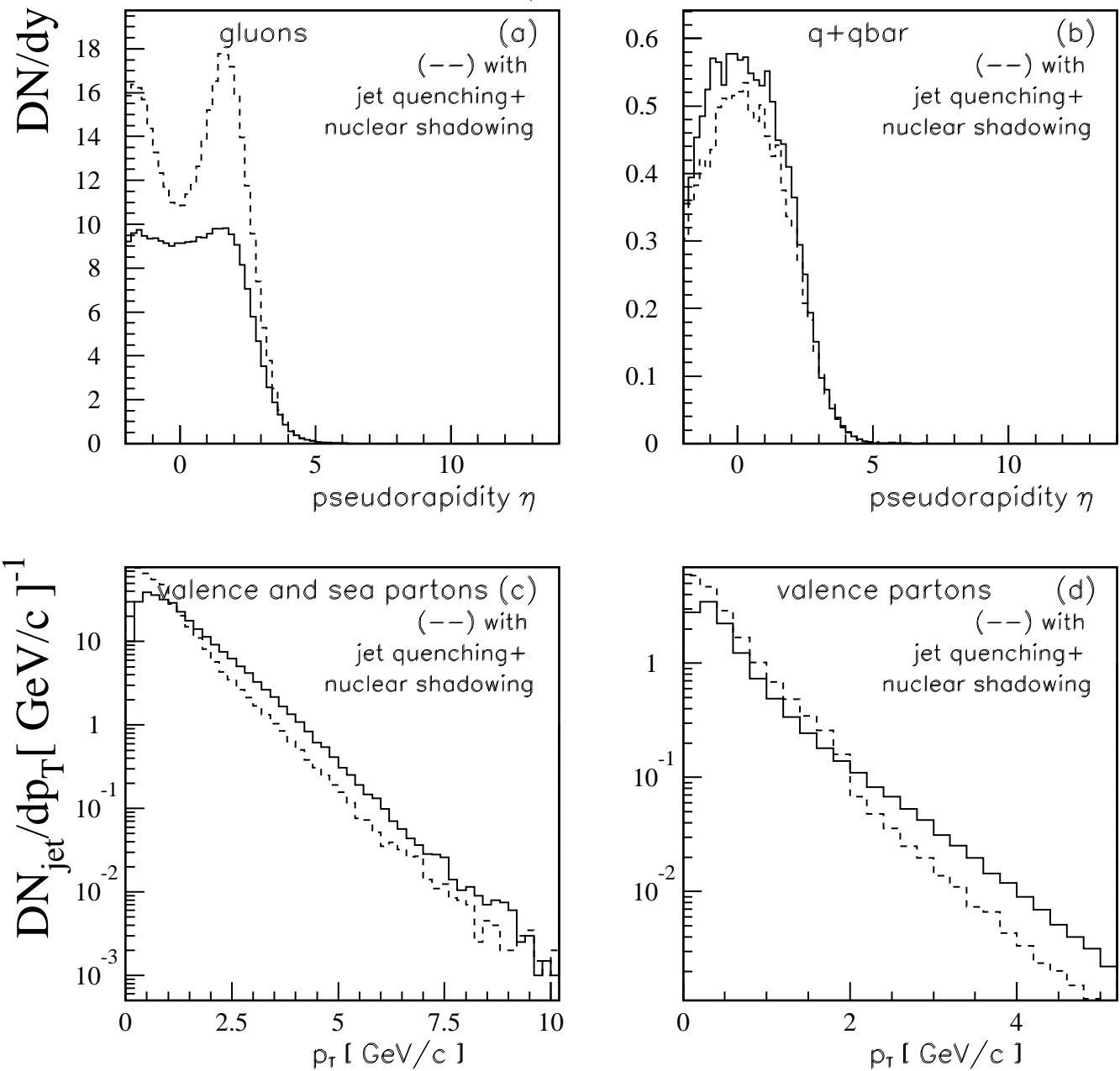
\begin{figure}
\caption{Rapidity distributions for gluons (Fig.19a)
  and quark-antiquark pairs ($q\bar{q}$)  (Fig.19b) ,
  transverse momenta distributions for valence and sea partons
  (Fig.19c) and valence parton (Fig.19d)  for
  Fe+Air interactions at 17.86 TeV/Nucleon and impact parameter
  interval (0-5 fm).The solid and dashed histograms have the same
   meaning as in Figure 18.}
\end{figure}

\newpage
\begin{table}
\caption{Particle multiplicities for $pp$ interaction
  at $200\, GeV\,$ are compared with data
   from Gazdzicki and Hansen \protect{\cite{1na35}} and
  with the results given by VENUS model (as computed here) and
  DPM model  DPMJET II \protect{\cite{ranf94a}}.}
\label{Tab1}
\vskip 0.5cm
\begin{tabular}{||c|c|c|c|c|c||}    \hline \hline
 {\bf pp} & {\bf Exp.data}  & {\bf HIJING} &${\bf HIJING^{(j)}}$
  &{\bf VENUS} &{\bf DPMJET II}\\
\hline
\hline
  $ <\pi^{-}>$  & $2.62\pm 0.06$ & $2.61$ & $2.65$  &$2.60$ & $2.56$\\
 \hline
   $ <\pi^{+}>$ & $3.22\pm 0.12$  & $3.18$ & $3.23$  & 3.10  & $3.17$\\
 \hline
   $<\pi^{0}>$  & $3.34 \pm 0.24$ & $3.27$ & $3.27$ &3.28 & $3.38$ \\
 \hline
   $<h^{-}>$  & $2.86 \pm 0.05$ & $2.99$ &$ 3.03$ & 3.05   & $2.82$ \\
 \hline
   $<K^{+}>$  & $0.28 \pm 0.06$ & $0.32$ & $0.32$  & 0.27 & $ 0.28$ \\
 \hline
   $<K^{-}>$  & $0.18 \pm 0.05$ & $0.24$& $0.25$  & 0.19 & $0.19$ \\
 \hline
  $<\Lambda + \Sigma^{0}>$ & $0.096 \pm 0.015$
  & $0.16$& $0.165$ & 0.18   &    \\
 \hline
   $<\bar{\Lambda}+\bar{\Sigma^{0}}>$  &
   $0.013 \pm 0.01$  & $0.03$ &
   $0.037$  &0.033 &    \\
 \hline
   $<K_{s}^{0}>$  & $0.17 \pm 0.01$ &
   $ 0.26$& $0.27$  & 0.27  & $0.22$  \\
 \hline
   $<p>$  & $ 1.34\pm 0.15 $  &$1.43$ & $1.45$ & 1.35 & $1.34$ \\
 \hline
   $<\bar{p}>$  & $0.05 \pm 0.02$ & $0.11$& $0.12$  &$ 0.06 $  & $0.07$ \\
 \hline
 \hline
\end{tabular}
\end{table}

\begin{table}
 \caption{Particle composition of $p+\bar{p}$ interactions
    at 540 GeV in cms.}
\label{Tab2}
\vskip 0.5cm
\begin{tabular}{||c||c|c|c||}
\hline\hline
{\bf Particle type} & ${\bf <n>}$  & $ {\bf Exp.data}$
& $ {\bf HIJING^{(j)}}$  \\
\hline
\hline
{\bf All charged} &  $29.4 \pm 0.3$ &\cite{1ua5} &  $28.2$  \\
\hline
 ${\bf K^{0}+\bar{K^{0}}}$ & $2.24 \pm 0.16$ & \cite{1ua5} & $1.98$ \\
\hline
 ${\bf K^{+}+K^{-}}$ &$ 2.24 \pm 0.16 $ & \cite{1ua5} & $2.06$ \\
\hline
 ${\bf p+\bar{p}}$ & $1.45 \pm 0.15$  & \cite{ward} &  $1.55$ \\
\hline
 ${\bf \Lambda+\bar{\Lambda}}$ & $0.53 \pm 0.11$
 & \cite{1ua5} & $0.50$ \\
\hline
  ${\bf \Sigma^{+}+\Sigma^{-}+\bar{\Sigma^{+}}+\bar{\Sigma^{-}}}$  &
  $0.27 \pm 0.06$  &    \cite{ward}    & $0.23$  \\
\hline
   ${\bf \Xi^{-}}$  &  $0.04 \pm 0.01$ & \cite{1ua5} & $0.037$  \\
\hline
  ${\bf \gamma}$  & $33 \pm 3$  & \cite{1ua5} &  $29.02$  \\
\hline
   ${\bf \pi^{+}+\pi^{-}}$  &  $23.9 \pm 0.4$ &\cite{1ua5} &   $23.29$ \\
\hline
   ${\bf K_{s}^{0}}$ &  $1.1 \pm 0.1$& \cite{1ua5}  &   $0.99$  \\
\hline
   ${\bf \pi^{0}}$ & $11.0\pm 0.4$  & \cite{ward} &    $13.36$ \\
\hline
\hline
\end{tabular}
\end{table}

\begin{table}
\caption{Percentage of occurence of various particles among
 the secondaries of a Nucleus-Air collision as calculeted
 by the HIJING and VENUS models.The number of
 protons and neutrons have been reduced by respective
 numbers in the primary system(values labeled by star(*)
 \protect{\cite{schatz1}}).
 The average multiplicity and numbers of participants are
  also given .}
\label{Tab3}
\vskip 0.5cm
\begin{tabular}{||c||c|c|c|c|c|c|c||}  \hline \hline
  ${\bf Particle}$ &${\bf Projectile}$ &${ \bf  {^5}{^6}Fe}$ &
  $ { \bf {^3}{^2}S}$ & ${ \bf {^2}{^0}Ne}$
  &${ \bf {^4}He }$ & ${\bf p}$ & ${\bf p}$\\
   ${\bf type} $ & ${\bf Energy(TeV/N)}$ &$ 17.86$ & $ 17.86$
  &$ 17.86$&$ 17.86 $ & $17.86$ & $1000$ \\
\hline
\hline
      ${\bf <\pi^-+\pi^+>}$ &$ {\bf HIJING}$ & $ 45.98$ &$45.93$ &
      $ 45.96$ & $45.86$  & $ 45.76$ & $46.57$ \\
      &${\bf VENUS}$ &$ 51.02$ &$ 51.29$ &$51.48$ &$52.34 $
       & $53.05$ & $52.15$ \\
\hline
\hline
  ${\bf <\pi^0>}$ & $ {\bf HIJING}$ & $26.13$ & $26.02$ & $26.10$ &
   $25.93$ & $26.04$ & $26.38$ \\
   &${\bf VENUS}$ &$28.30$ & $28.52$ & $28.49$ & $28.85$ & $28.93$ &
   $28.43$ \\
\hline
\hline
   ${\bf <K^++K^->}$ & $ {\bf HIJING}$ & $5.20$ & $5.20$ & $5.20$ &
   $5.20$ & $5.16$ & $5.58$ \\
   $ {\bf K \,\,mesons}  $ &${\bf VENUS}$ &$12.35$ & 12.02 & 11.84 &
    11.11 &  10.51 & 10.86 \\
\hline
\hline
    ${\bf <K^0_s>}$ & $ {\bf HIJING}$ & $2.54$ &$2.57$ & $2.53$ &
    $2.52$ & $2.57$ & $2.75$  \\
	    & ${\bf VENUS}$ &   &   &   &  &  &   \\
\hline
\hline
  ${\bf <p> }$ & $ {\bf HIJING}$ & $4.16$ & $4.20$ & $4.20$ &
   $4.37$ & $4.78$ & $3.25$  \\
   ${\bf <p> *}$ & ${\bf VENUS}$ & $0.70$ & $0.66$ & $0.66$ &
   $0.66$ & $-0.11$ & $0.66$ \\
\hline
\hline
 ${\bf <n> }$ & $ {\bf HIJING}$ & $4.12$ & $4.20$ & $4.23$ &
  $4.40$ & $4.16$ & $2.86$  \\
  ${\bf <n> *}$ & ${\bf VENUS}$ & $0.54$ & $0.61$ & $0.64$ &
  $0.66$ & $1.49$ &$ 1.43$  \\
\hline
\hline
 ${\bf <\Lambda>}$ & $ {\bf HIJING}$ & $0.70$ & $0.70$ &$0.71$ &
 $0.72$ & $0.72$ & $0.56$ \\
 ${\bf <\Lambda+\Sigma^0>}$ & ${\bf VENUS}$
 & $1.66$ & $1.64$ &$1.67$ &
  $1.60$ & $1.55$ & $1.31$  \\
\hline
\hline
 ${\bf Other\,\, baryons}$ &$ {\bf HIJING}$ &   &  &  &  &  &   \\
	&${\bf VENUS}$ & $0.46$ & $0.43$ & $0.42$ & $0.34$ & $0.26$
	 & $0.28$ \\
\hline
\hline
 ${\bf <\gamma>}$ & ${\bf HIJING}$ & $4.30$ & $4.30$ & $4.33$ &
  $4.23$ & $4.27$ & $4.48$    \\
	&${\bf VENUS}$ & $1.60$ & $1.48$ & $1.43$ & $1.20$ &
	$1.12$  & $1.18$  \\
\hline
\hline
 ${\bf all\,\, charged}$ & $ {\bf HIJING}$ & $57.30$
 & $57.47$ & $57.54$ &$57.54$ & $57.84$ & $57.79$ \\
\hline
\hline
 ${\bf all\,\, neutrals}$ & ${\bf HIJING}$ & $42.33$
 & $42.57$ & $42.61$ &
    $42.50$ & $42.40$ & $42.18$  \\
\hline
\hline
 ${\bf <\bar{p}>}$ & ${\bf HIJING}$ & $1.42$ & $1.38$ & $1.43$ &
 $1.40$ & $1.41$ & $1.56$  \\
\hline
\hline
${\bf <\bar{n}>}$ & ${\bf HIJING}$ & $1.40$ & $1.38$ & $1.43$ &
 $1.40$ & $1.41$ & $1.60$ \\
\hline
\hline
 ${\bf <gluons>}$ & ${\bf HIJING}$ & $9.70$ & $9.20$ & $8.96$ &
 $7.73$ & $7.20$ & $10.71$ \\
\hline
\hline
 ${\bf <q+\bar{q}>}$ & ${\bf HIJING}$ & $0.40$ & $0.43$ & $0.41$ &
 $0.36$ & $0.33$ & $0.71$ \\
\hline
\hline
 ${\bf Mean}$ &${\bf HIJING}$ & $274.0$ & $217.5$ & $203.0$ &
 $88.7$ & $38.9$ & $51.7$  \\
 ${\bf multiplicity}$ & ${\bf VENUS}$ & $354.7$ & $270.5$ &
 $209.3$ & $92.6$ & $49.4$ & $106.5$ \\
\hline
\hline
 ${\bf Mean\,\, projectile}$ &${\bf HIJING}$ & $9.6$ & $7.2$ & $6.5$ &
 $2.2$ & $1.0$ & $1.0$ \\
 ${\bf participants}$ & ${\bf VENUS}$ & $12.1$ & $8.1$ & $5.7$ &
 $2.1$ & $1.0$ & $1.0$ \\
 \hline
 \hline
 ${\bf Mean\,\, target}$ & ${\bf HIJING}$ & $5.9$ & $5.6$ & $5.6$ &
   $3.4$ & $2.06$ & $1.42$ \\
 ${\bf participants}$ & ${\bf VENUS}$ & $6.0$ & $5.3$ & $4.7$ &
  $3.2$ & $2.0$ & $2.06$ \\
\hline
\hline
\end{tabular}
\end{table}

\begin{table}
\caption{Mean transverse energy for
 the secondaries of a Nucleus-Air collisions
 at 17.86 TeV/Nucleon as calculeted
 by the HIJING  model and by considering superposition
 of $N_{coll}$ nucleon-nucleus collisions and $N_{proj}$
 nucleon-nucleus collisions,where $N_{coll}$ is the number
 of binary collisions and $N_{proj}$ is the number of
 participant projectile nucleons. $E_{T}^{pA}$ is
 mean transverse energy in p+Air interaction at the same
 energy and in the same impact parameter interval
 (see the text for explanation).}
\label{Tab4}
\vskip 0.5cm
\begin{tabular}{||c||c|c|c|c|c|c||}  \hline \hline
        &${\bf Projectile}$ &${ \bf  {^5}{^6}Fe}$ &
  $ { \bf {^3}{^2}S}$ & ${ \bf {^2}{^0}Ne}$
  &${ \bf {^4}He }$ & ${\bf p}$  \\
\hline
\hline
      ${\bf Mean \,\,number}$ & $<N_{coll}>$ & $27.75$ & $19.21$ &
      $13.61$ & $2.62$ &   \\
\hline
\hline
 ${\bf Mean\,\, number}$   & $<N_{proj}>$ & $18.43$ & $11.44$ &
	$8.24$ & $1.73$ &  \\
\hline
\hline
      ${\bf  Mean\,\, Transverse}$ &$ {\bf all\,\, secondaries}$ &
   $212.5$ & $ 140.7$ & $104.2$  & $ 25.16$ & $7.54$ \\
       $\bf Energy$ &${\bf N_{coll}*E_{T}^{pA}}$ &
     $ 209.2$ &$ 144.8$ &$102.6$ &$19.75 $ &    \\
       $ {\bf (GeV)}$   &${\bf N_{proj}*E_{T}^{pA}}$ &
     $138.96$ & $86.25$ & $ 62.10$ & $13.04$ &  \\
\hline
\hline
 ${\bf  Mean\,\, Transverse}$ &$ {\bf all\,\, charged} $ &
 $122.7$ & $81.27$ & $60.10$ & $14.52$ & $4.39$ \\
 $\bf Energy$ &${\bf N_{coll}*E_{T}^{pA}}$ &
   $121.8$ & $84.33$ & $59.74$ & $11.50$ &  \\
  $ {\bf (GeV)}$   &${\bf N_{proj}*E_{T}^{pA}}$ &
  $80.90$ & $50.22$ & $36.17$ & $7.6$ &  \\
\hline
\hline
 ${\bf  Mean\,\, Transverse}$ &$ {\bf all\,\,neutrals}$ &
 $89.87$ & 59.43 & $44.10$ & $10.64$ & $3.15$ \\
 $\bf Energy$ &${\bf N_{coll}*E_{T}^{pA}}$ &
 $87.41$ & $60.51$ & $42.87$ & $8.25$ &  \\
  $ {\bf (GeV)}$   &${\bf N_{proj}*E_{T}^{pA}}$ &
  $58.05$ & $36.03$ & $25.95$ & $5.44$ & \\
\hline
\hline
\end{tabular}
\end{table}


\newpage
\begin{thebibliography}{199}
\bibitem{rebel1} H.Rebel in {\it Topics in Atomic and Nuclear
Collisions (NATO ASI Series }{\bf B321},eds. B.Remaud ,
 A.~Calboreanu and V.~Zoran (New Plenum),p397, (1994)
\bibitem{mul1} D.Muller,S.P.Swordy,P.Meyer,L.Heureux and J.M.
Grunsfeld , Astrophys.J.{\bf 374 } (1991) 356\\
 S.P.Swordy,L.Heureux,P.Meyer and D.Muller,
Astrophys.J.{\bf 403} (1993) 658
\bibitem{rebel2} A.Haungs,J.~Kempa,H.J.~Mathes,H.~Rebel,J.~Wentz\\
Preprint submitted to Elsevier Science (July 1995)
\bibitem{topor2}  M.Gyulassy and V.Topor Pop  ,Preperint CU- 95
(in preparation)
\bibitem{ranf94a} J.Ranft,Preprint LNF-94/035(P),Laboratori
Nazionali di Frascati (Submitted to Phys.Rev.D)
\bibitem{ranf94b} G.Battistoni,C.Forti,J.Ranft, Preprint LNF-94/048,
Laboratori Nazionali di Frascati ; Astroparticle Physics
{\bf 3} (1995) 157
\bibitem{dpm94} A.Capella,U.Sukhatme,C.~I.~Tan and
J.~Tran ~Thanh ~Van ,Phys.~Rep.{\bf 236},225(1994)
\bibitem{ame1}N.~S.~Amelin ,E.F.Staubo,L.P.Csernai,V.D.Toneev,K.K.Gudima, \\
Phys.Rev.{\bf C44},1541(1991)\\
N.~S.~Amelin,L.V.Bravina,L.P.Csernai,V.D.Toneev
K.K.Gudima,S.Yu. Sivoklokov,\\
Phys.Rev.{\bf C47},2299(1993)
\bibitem{wer7}K.~Werner,Preprint HD-TVP-93-1(1993),University of
 Heidelberg,Germany;\\
 Phys.Rep.{\bf 232},87(1993)
\bibitem{ander}B.Andersson,G.Gustafson and
Nilsson-Almqvist Nucl.Phys.{\bf B281},289(1987);\\
B.Nilsson-Almqvistand E.Stenlund,
Comp.Phys.Commun.{\bf 43},387(1987)
\bibitem{wang0}Xin-Nian Wang and Miklos Gyulassy,
Comp.Phys.Comm.{\bf 83},307(1994)
\bibitem{wang1}Xin-Nian Wang,Phys.Rev.{\bf D43},104(1991)
\bibitem{wang2}Xin-Nian Wang and Miklos Gyulassy,Phys.Rev.
{\bf D44},3501(1991)
\bibitem{wang3}Xin-Nian Wang and Miklos Gyulassy,Phys.Rev.
{\bf D45},844(1992)
\bibitem{geiger95} K.Geiger,Phys.Rep. {\bf 258}, 237 (1995)
\bibitem{amelin95} N.S.Amelin,H.Stocker,W.Greiner,
 N.~Armesto, \\
  M.A.~Braun and C.~Pajares,Phys.Rev. {\bf C52},362, (1995)
\bibitem{capde89} J.~N.~Capdevielle ,J.~Phys.~{\bf G}:Nucl.Phys.
{\bf 15} (1989) 909
\bibitem{forti90}C.Forti,H.~Biloken,B.Piazzoli,T.K.Gaisser,
L.~Satta and T.~Stanev, Phys.Rev.{\bf D42} (1990) 3668
\bibitem{flet94} R.S.Fletcher,T.K.~Gaisser,P.~Lipari and
T.Stanev,Bartol Preprint BA 94-01,submitted to
Phys.~Rev.~ D (1994)
 \bibitem{schatz1} G.~Schatz,T.~Thouw,K.~Werner,J.~Oehlschlager and
 K.~Bekk ,J.~Phys.~G:\\
 ~Nucl.~Part.~Phys.{\bf 20}(1994)1267
 \bibitem{capde92} J.~N.~ Capdevielle et al.,  KFK Report
 {\bf 4998},Kernforschungzentrum Karlsruhe (1992)
 \bibitem{topor1} V.Topor Pop,M.Gyulassy,X.N.Wang,A.Andrighetto,
 M.Morando,F.Pellegrini,\\
 R.A.Ricci,G.Segato ,
 Preprint CU-676(1995),Columbia University,New York,\\
 Phys.Rev.C (in print)
\bibitem{sjos94}T.Sjostrand,Comp.Phys.Comm.{\bf 82},74(1994)
\bibitem{gaisser} A.~Capella and J.~Tran Thanh Van, Z.~Phys. {\bf C
23},165 (1984);  T.~K.~Gaisser and F.~Halzen,
 Phys. Rev. Lett. {\bf 54},1754 (1985);
 G.~Pancheri and Y.~N.~Srivastava, Phys. Lett. {\bf 182B},
199 (1986); L.~Durand and H.~Pi, Phys. Rev. Lett. {\bf 58}, 303
(1987)
\bibitem{kaja}K.~J.~Eskola, K.~Kajantie and J.~Lindfors,
Nucl. Phys. {\bf B323}, 37 (1989)
\bibitem{eichten} E.~Eichten, I.~Hinchliffe and C.~Quigg, Rev. Mod.
Phys. {\bf 56}, 579 (1984).
\bibitem{duke} D.~W.~Duke and J.~F.~Owens, Phys. Rev. {\bf D 30}, 50
(1984).
\bibitem{shadow1} EM Collab., J.~Ashman, {\em et al.}, Phys. Lett.
{\bf 202B},
603 (1988); EM Collab., M.~Arneodo, {\em et al.}, Phys. Lett.
{\bf 211B}, 493 (1988).
\bibitem{shadow2} A.~H.~Mueller and J.~Qiu, Nucl. Phys. {\bf B 268},
427 (1986); J.~Qiu, Nucl. Phys. {\bf B 291}, 746 (1987).
\bibitem{dedx} M.~Gyulassy and X.~N.~Wang, preprint LBL-32682.
\bibitem{knapp2} H.H.Mielke,M.Foller,J.Engler and J.Knapp,
 J.Phys.{\bf G};\\
 Nucl.Part.Phys. {\bf 20}(1994)637
 \bibitem{1na35} M.Gazdzicki and O.Hansen,Nucl.Phys.{\bf A528}(1991),754
 \bibitem{ex80} F.Lo Pinto et al., Phys.Rev. {\bf D22}, 573 (1980)
\bibitem{1ua5} G.J.Alner et al.,Phys.Rep.{\bf 154},247(1987)
\bibitem{ward} D.~R.~Ward {\it Properties of Soft
Proton Antiproton Collisions }in Advances Series on
Direction in High Energy Physics ,Vol.4 ,Eds.G.Altarelli and
L.Di Lella,1989 ,p.85
\bibitem{drem} I.~M.~Dremin ,Sov.J.Nucl.Phys.
{\bf 33},726(1981)
\bibitem{mik95} M.Gyulassy in
Proc.of Eleventh International Conference on
Ultrarelativistic Nucleus-Nucleus Collisions,
Quark Matter '95 (Monterey,California,USA 9-13 January,1995),
edited by  A.M.Poskanzer, J.W.Harris and L.S.Schroder
Nucl.Phys.{\bf A590 } , 431c (1995)
\bibitem{alber94} T.Alber et al.,Z.f\"{u}r Physik {\bf C64}(1994)195
\bibitem{baec93} J.Baechler et al.,Z.f\"{u}r Physik {\bf C58}
(1993) 367
\bibitem{murray95} M.Murray et al.,
Proc.of  International Workshop   {\it Strangeness in Hadronic Matter},
 January 4-7 ,1995,Tucson,Arizona (to be published)
  (private communication)
\bibitem{awes89} T.C.Awes,S.P.Sorensen , Nucl.Phys.{\bf A498},
123c(1989).
\end{thebibliography}
\end{document}